%% file: PoS_2022_draft.tex
\documentclass[a4paper,11pt]{article}
\usepackage{pos}
\usepackage{color}
\usepackage{tikz}
\usepackage{cancel}
\usetikzlibrary{calc}
\def\lVt{l_{\rm \widetilde{V}}}
\def\gVt{g_{\rm \widetilde{V}}}

\input{macros.tex}

\title{RG-running of the tensor currents for $N_f$ =3 QCD in a $\chi SF$ setup}

\author[a]{\begin{center}
     \includegraphics[height=2.0\baselineskip]{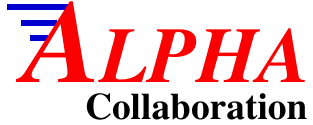}
\end{center} Isabel Campos Plasencia}
\author[b]{Mattia Dalla Brida}
\author[c,d]{Giulia Maria de Divitiis}
\author[e]{Andrew Lytle}
\author[f,g]{Mauro Papinutto}
\author*[c,d]{Ludovica Pirelli}
\author[d]{Anastassios Vladikas}

\affiliation[a]{Instituto de F\'isica de Cantabria IFCA-CSIC,\\
  Avda. de los Castros s/n, 39005, Santander, Spain}
 \affiliation[b]{Theoretical Physics Department, CERN, \\
 CH-1211, Geneva 23, Switzerland}
 \affiliation[c]{Dipartimento di Fisica, Universit\`a di Roma ``Tor Vergata'', \\
 Via della Ricerca Scientifica 1, 00133 Roma, Italy}
 \affiliation[d]{INFN, Sezione di Tor Vergata, \\
 Via della Ricerca Scientifica 1, 00133 Roma, Italy}
 \affiliation[e]{Department of Physics, University of Illinois at Urbana-Champaign, \\
 Urbana, Illinois, 61801, USA}
 \affiliation[f]{Dipartimento di Fisica,  Universit\`a di Roma La Sapienza, \\
 Piazzale A.~Moro 2, 00185 Roma, Italy}
\affiliation[g]{INFN, Sezione di Roma, \\
Piazzale A.~Moro 2, 00185 Roma, Italy}

\emailAdd{isabel.campos@csic.es}
\emailAdd{mattia.dalla.brida@cern.ch}
\emailAdd{giulia.dedivitiis@roma2.infn.it}
\emailAdd{atlytle@illinois.edu}
\emailAdd{mauro.papinutto@roma1.infn.it}
\emailAdd{ludovica.pirelli@roma2.infn.it}
\emailAdd{tassos.vladikas@roma2.infn.it}

\abstract{
We present the preliminary results of the non-perturbative running of the flavour non-singlet tensor operator in the high-energy range  $2~\rm{GeV}\lesssim \mu\lesssim 128~\rm{GeV}$ in  $\NF=3$ massless QCD, comparing  four different definitions of the renormalisation constant.  We use the configuration ensembles of ref.\cite{Campos:2018ahf} and ref.\cite{DallaBrida:2016kgh}, subject to Schr\"odinger functional (SF) boundary conditions, and valence quarks with chirally rotated Schr\"odinger functional ($\chi$SF) boundary conditions.  Provided that boundary counterterms have been appropriately tuned, this results in O($a$)  improvement of the tensor operator, without the need of a dimension-4 Symanzik counterterm (proportional to $c_T$).
}

\FullConference{%
  The 39th International Symposium on Lattice Field Theory (Lattice2022),\\
  8-13 August, 2022 \\
  Bonn, Germany 
}

\begin{document}
\begin{flushright}
CERN-TH-2022-185
\end{flushright}
\maketitle

\section{Flavour non-singlet tensor operator}
A non-perturbative determination of renormalisation group running between hadronic and electroweak scales  for the flavour non-singlet tensor operator
\begin{align}
T_{\mu \nu} ^{f_1 \, f_2}(x)=i\bar \psi_{f_1} (x) \; &\sigma_{\mu \nu} \; \tfrac{1}{2} \; \psi_{f_2}(x)
\end{align} 
is very interesting from both phenomenological and theoretical points of view.
The tensor enters the amplitudes of effective Hamiltonians describing possible Beyond Standard Model effects, for example, in rare heavy meson decays (see for example ref. \cite{Blake:2016olu}) or neutron beta decays (see e.g. ref.\cite{Bhattacharya:2011qm}).
Moreover, the computation of the scale dependence of the renormalisation factor completes the ALPHA renormalisation and improvement programme of the dimension-3 bilinear operators.
For $N_f=0,2$ such a study has appeared in ref.\cite{Pena:2017hct}. For $N_f=3$, preliminary results of the RG-running in the high 
energy range $2~\rm{GeV}\lesssim \mu\lesssim 128~\rm{GeV}$ have been reported in ref.\cite{Chimirri:2019xsv}. 
These results were obtained in a Schr\"odinger functional (SF) setup, while the ones in the presented work are obtained with chirally rotated Schr\"odinger functional ($\chi$SF) boundaries for the valence quarks. The two setups have the same continuum limit, but $\chi$SF also benefits from automatic $O(a)$ improvement. (refs.\cite{Sint:2010eh,Sint:2010xy,DallaBrida:2016smt,DallaBrida:2018tpn,Mainar:2016uwb}).

\section{RG flow}
In a mass-independent renormalisation scheme we can define the following RG equation for the renormalised operator $T_\mathrm{R}(\mu) = \ZT(\mu) T $: %
\begin{align}
  \mu \frac{\partial}{\partial \mu} T_\mathrm{R}(\mu) =\gamma  (g_\mathrm{R}(\mu)) \, &T_\mathrm{R}(\mu)\,,\qquad\qquad  
  \end{align}
where $g_\mathrm{R}$ is the running coupling.
The anomalous dimension $\gamma$   has the perturbative expansion
\begin{align}
  \gamma(g_\mathrm{R})  &\stackrel{g_\mathrm{R}\to0}{\sim} -{g_\mathrm{R}}^2( \gamma_0 +\gamma_1 {g_\mathrm{R}}^2 +\gamma_2 {g_\mathrm{R}}^4 + \mathcal{O}({g_\mathrm{R}}^6)) \,,
\end{align}

where $\gamma_0$ is a universal coefficient.
From a particular solution of the RG equation we can extract the corresponding renormalisation group invariant (RGI):

\begin{align}
       {T}_{\scriptscriptstyle \rm  RGI} & =  {{T} _\mathrm{R}(\mu)} \left [ \frac{ {g_\mathrm{R}}^2(\mu) }{4 \pi} \right ]^{-\frac{\gamma_0}{2b_0}}\exp  \left \{
            -\int\limits_0^{g_\mathrm{R}(\mu)} \, { d\mathnormal{g}} \,{\left [
            \frac{\gamma(\mathnormal{g})}{\beta(\mathnormal{g})}-\frac{\gamma_0}{b_0 \mathnormal{g}} \right ]} \right \} \,.
\end{align}

It is possible to factorise the running in many evolutions between two scales:
\begin{align}{
{ {T}_{\scriptscriptstyle \rm  RGI}}= 
{\frac{T_{\scriptscriptstyle \rm  RGI}}{T_\mathrm{R}(\mu_{pt})}}\;
\cdots
{\frac{T_\mathrm{R}(\mu_{1})}{T_\mathrm{R}(\mu_2)}}\;
{\frac{T_\mathrm{R}(\mu_2)}{T_\mathrm{R}(\mu_3)}}\;
{\frac{T_\mathrm{R}(\mu_3)}{T_\mathrm{R}(\mu_{\rm{had}})}}\;
T_\mathrm{R}(\mu_{\rm{had}})
\,,
}\end{align}
leading naturally to the definition of the step scaling function: 
\begin{align}
\sigma_{T}(s,u)=
\frac{T_\mathrm{R}(\mu_2)}{T_\mathrm{R}(\mu_1)}=
\frac{Z_\mathrm{T}(\mu_2)}{Z_\mathrm{T}(\mu_1)}\,,
\end{align}
where $s\equiv\frac{\mu_1}{\mu_2}$ and $u\equiv g^2_\mathrm{R}(\mu_1)$.
A common and convenient choice is to take successive scales at fixed ratio $s=2$:
\begin{align}
\label{eq:sigmaT_def}
\sigma_{T}(u)\equiv\sigma_{T}(2,u)=
      \exp  \left \{
            \int\limits_{g_\mathrm{R}(\mu)}^{g_\mathrm{R}(\mu/2)} \, { d\mathnormal{g}} \,{
            \frac{\gamma(\mathnormal{g})}{\beta(\mathnormal{g})}} \right \} \,.
\end{align}
On the lattice, the  scale evolution can be studied non-perturbatively as a finite size scaling,
with the renormalisation scale identified as the inverse of the lattice size $\mu = \frac{1}{L}$:
\begin{alignat}{3}
&{\mu = \frac{1}{L}}, \quad u\equiv {g_\mathrm{R}}^2(L)&\qquad&\\
&\sigma(u)=\lim_{a\to 0} \Sigma(u,a/L)   && \Sigma(u,a/L)={g_\mathrm{R}}^2(2L)\\
&\sigma_{T}(u)=\lim_{a\to 0} \Sigma_{T}(u,a/L) && \Sigma_{T}(u,a/L)=\frac{Z_{T}(g_0^2,a/2L)}{Z_{T}(g_0^2,a/L)}\,,
\end{alignat}
where $a$ is the lattice spacing.
The renormalisation constants $Z_{T}(g_0^2,a/L)$  are defined imposing renormalisation conditions on the correlation functions, as shown in Eq.\eqref{eq:def-Zt} of section \ref{cond_renorm}.

\section{Computational setup}
We used the same gauge configurations generated  by the ALPHA collaboration for the determination of the quark mass running (see  ref.\cite{Campos:2018ahf} and ref.\cite{DallaBrida:2016kgh} for details of the simulations).
They refer to $N_f=3$ massless Wilson-clover fermions with Schr\"odinger Functional (SF) boundary conditions. However, here we work in a mixed action setup (see also ref.\cite{Plasencia:2021eon}): sea quarks are regularised in SF, valence quarks in $\chi$SF.
The simulation parameters correspond to a RG evolution from 
 an hadronic scale $\mu_{had}$ of about $200~\rm{MeV}$ to a perturbative scale $\mu_{pt}$ around $128~\rm{GeV}$ (see refs.\cite{Bruno:2017gxd},\cite{DallaBrida:2016uha}). 
The peculiarity  of this RG flow is the change of schemes at the intermediate scale 
$\mu_0/2 \sim 2~\rm{GeV}$: in the high energy region the running coupling is defined in the SF scheme $(g_\mathrm{R}={g}_{SF})$ (ref.\cite{Luscher:1992an}),
while in the low energy region it is defined in the gradient flow (GF) scheme $(g_\mathrm{R}={g}_{GF})$ (ref.\cite{Fritzsch:2013je}):
\begin{center}
 \begin{tikzpicture}
 \coordinate (C);
\draw[->, line width=0.5mm, black]  ($(C)+(0.0,-2.0)$) to node {}($(C)+(8.0,-2.0)$);
\node at ($(C)+(5.5,-1.8)$) [above] {SF scheme};
\node at ($(C)+(2.5,-1.8)$) [above] {GF scheme};
\node at ($(C)+(8.5,-2.0)$) [left] {$\mu$};
\draw[line width=0.5mm, black]  ($(C)+(1.0,-1.9)$) to node[above] {$\mu_{had}$}($(C)+(1.0,-2.1)$);
\draw[line width=0.5mm, black]  ($(C)+(1.0,-1.9)$) to node[below] {$\approx \, 200 \mathrm{MeV}$}($(C)+(1.0,-2.1)$);
\draw[line width=0.5mm, black]  ($(C)+(7.0,-1.9)$) to node[above] {$\mu_{pt}$} ($(C)+(7.0,-2.1)$);
\draw[line width=0.5mm, black]  ($(C)+(7.0,-1.9)$) to node[below] {$\approx \, 128 \mathrm{GeV}$} ($(C)+(7.0,-2.1)$);
\draw[line width=0.5mm, black]  ($(C)+(4.0,-1.9)$) to node[above] {$\mu_0/2$} ($(C)+(4.0,-2.1)$);
\draw[line width=0.5mm, black]  ($(C)+(4.0,-1.9)$) to node[below] {$\approx \, 2 \mathrm{GeV}$} ($(C)+(4.0,-2.1)$);
\end{tikzpicture}
\end{center}

We impose the same definition of $\ZT(g_0^2,a/L)$  at all scales, which implies that the anomalous dimension has the same value at a given renormalisation scale $\mu$:
\begin{align}
\gamma(\mu)=\gamma_{SF}(g^2_{SF}(\mu))=\gamma_{GF}(g^2_{GF}(\mu))\,.
\end{align}

\section{Renormalisation schemes in $\chi$SF}
\label{cond_renorm}
At a formal level, continuum massless QCD with $\chi$SF boundary conditions is obtained from its SF counterpart by a chiral non-singlet transformation of the fermion fields (ref.\cite{Sint:2010eh}): 
\begin{equation}
\label{eq:ferm-rots}
\psi = R(\pi/2) \, \psi^\prime \,\, , \qquad  \bar \psi = \bar \psi^\prime \,  R(\pi/2) \,\, ,
\end{equation}
where  $\psi, \bar \psi$ and $\psi^\prime, \bar \psi^\prime$ are doublets in isospin space and $R(\alpha) = \exp(i \alpha \gamma_5 \tau^3/2)$. We can map SF correlation functions into $\chi$SF ones. We point to ref.\cite{DallaBrida:2016smt} for the definitions. We only quote the continuum relations for the boundary-to-bulk correlation functions related to the tensor and vector current:
\begin{align}
\label{eq:bulk-bnd}
\kT=\lT^{ud} & \, & \fA=-i \gV^{ud} & \, & \kV= \lV^{uu'} \, ,
\end{align}
and for the boundary-to-boundary correlation functions
\begin{equation}
\label{eq:bnd-bnd}
\begin{array}{ccc}
f_1 = g_1^ {ud}  &\, ,&  k_1 = l_1^ {ud}   \ .
\end{array}
\end{equation}
These formal identities follow from the invariance of the massless QCD action under flavour and chiral transformations. They are broken on the lattice, but  they are recovered after renormalisation in the continuum limit.
The above correlation functions will be used in the definition of the renormalisation constant $\ZT$ in $\chi$SF, for a symmetric lattice with volume $L^3 \times T$  and for $T=L$.
Thanks to the property of automatic $O(a)$-improvement of  $\chi$SF, the tensor correlator $\lT^{ud}$ does not need the Symanzik correction (see also ref.\cite{deDivitiis:2021ugi})
  \begin{align}
&\qquad T_{\mu \nu}^{\rm I}=T_{\mu \nu} + { c_{\rm T}(g_0^2)\; a\,(\tilde{\partial}_{\mu}V_{\nu} - \tilde{\partial}_{\nu}V_{\mu})}\,,\\
&\qquad \lT^{{ud}, {\rm I}}=\lT^{ud} +\cancel{c_{\rm T}(g_0^2)\; a\,\tilde{\partial}_{0}\lV^{ud}} \,.
\end{align}
Following ref.\cite{Mainar:2016uwb} and ref.\cite{Pena:2017hct}, we have some freedom in the choice of the normalisation in the definition of $\ZT$. That, along with the parameter $\theta$ entering spatial boundary conditions (ref.\cite{DallaBrida:2016smt}), fixes the renormalisation scheme:
 \begin{align}
& \qquad \ZT(g_0,a/L)\frac{ \lT^{ud} (L/2)}{(g_1^{ud})^\alpha (l_1^{ud})^\beta (\gVt^{ud})^\gamma (\lVt^{uu'})^\delta } =
 \left.\frac{ {\lT^{ud} (L/2)}}{(g_1^{ud})^\alpha (l_1^{ud})^\beta (\gVt^{ud})^\gamma (\lVt^{uu'})^\delta }\right|^{\text{Tree Level}}  \, ,
 \label{eq:def-Zt}
\end{align}
with the condition
\begin{align}
& \alpha + \beta + \frac{1}{2} \gamma +  \frac{1}{2} \delta = 1 \, .
\label{eq:par-cond}
\end{align}
We are going to work with the renormalisation schemes defined by $\theta=0.5$, $T=L$ and 
\begin{equation}
(\alpha,\beta,\gamma,\delta)=\begin{cases}
(0.5,0,0,0) & \alpha{\rm -scheme}\\
(0,0.5,0,0) & \beta{\rm -scheme}\\
(0,0,1,0) & \gamma{\rm -scheme}\\
(0,0,0,1) & \delta{\rm -scheme}
\end{cases}
\label{eq:scheme_labels}
\end{equation}

The first two definitions in Eq. \eqref{eq:scheme_labels} are equivalent to the two SF ones of ref.\cite{Pena:2017hct} thanks to Eqs. \eqref{eq:bulk-bnd},\eqref{eq:bnd-bnd}. The last two definitions in Eq. \eqref{eq:scheme_labels} benefit from $Z_{\rm \widetilde{V}}=1$, since $\tilde{V}$ is the conserved lattice vector current (ref.\cite{DallaBrida:2016smt}). \\
It is possible to obtain the first non-universal coefficient $\gamma_1$ in the $\gamma(g_R)$ expansion for all the schemes, by relating them at one-loop order to a reference scheme where the two-loop value $\gamma_1$ is already known. 
The relation connecting two mass-independent schemes that differ only by the definition of the renormalised tensor operator is (see ref.\cite{Sint:1998iq} and ref.\cite{Pena:2017hct}):
\begin{align}
\gamma'(g)=\gamma(g)+\beta(g) \frac{\partial}{\partial g} \ln \chi(g)
\end{align}
where $\chi(g_R)$ is defined by
\begin{align}
T'_R = \chi(g_R) T_R \, ,
\label{eq:chi_T}
\end{align}
with perturbative expansion:
\begin{align}
\chi(g)= 1 + \chi^{(1)} g^2 + ...
\end{align}
In practice the connection between $\gamma_1$ of one $\chi$SF scheme and $\gamma_1$ of  $\overline{\mathrm MS}$ scheme reads :
\begin{align}
\gamma^{\chi SF}_1 = \gamma^{\overline{\mathrm MS}}_1 + 2 b_0 ( \chi^{(1)}_{\chi SF, \rm{lat}} - \chi^{(1)}_{\overline{\mathrm MS}} ) \, ,
\label{eq:chi_T_complete}
\end{align}
where $b_0$ is the first universal coefficient of the $\beta$-function, $\gamma^{\overline{\mathrm MS}}_1$ is taken from ref.\cite{Gracey:2000am}, $ \chi^{(1)}_{\overline{\mathrm MS}, \rm{lat}}$ from ref.\cite{Skouroupathis:2008mf} and $ \chi^{(1)}_{\chi SF, \rm{lat}} \equiv r_0$ is extracted from our fits to the perturbative results computed in ref.\cite{DallaBrida:2016smt} with the asymptotic parametrisation 
\begin{align}
Z_{T}= 1 + Z_{T}^{(1)} g^2 + ...   \, .
\end{align}

\begin{equation}
Z_{T}^{(1)}\sim \underset{n=0}{\overset{n_{max}}{\sum}}[r_{n}+s_{n}\ln(L/a)](\frac{a}{L})^{n} \, .
\label{eq:ZT1-exp}
\end{equation}
Further details will be given in ref.\cite{paper:2022}. We here write our preliminary values for $\gamma^{\chi SF}_1$:
\begin{equation}
\label{eq:gamma1}
\gamma^{\chi SF}_1=\begin{cases}
0.0062755(11) & \alpha-{\rm scheme}\\
0.0057956(11) & \beta-{\rm scheme}\\
-0.0007746(11) & \gamma-{\rm scheme}\\
0.0032320(11) & \delta-{\rm scheme} \, .
\end{cases}
\end{equation}

\section{Results}
\begin{figure}
\centering
\includegraphics[width=0.7\textwidth]{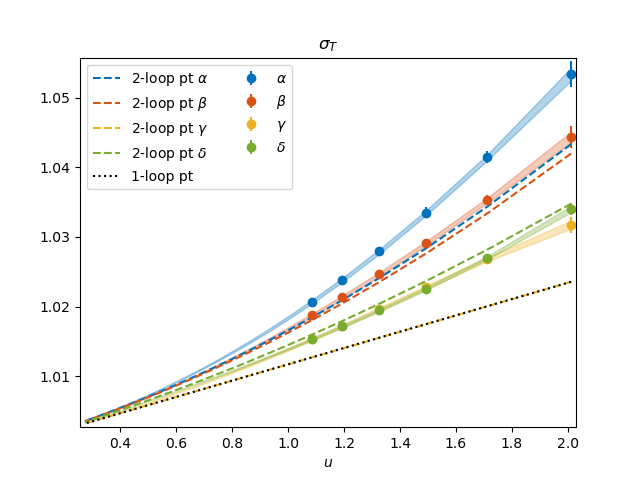}
\caption{$\sigma_{\rm T}(u)$ compared with the universal 1-loop and the 2-loop perturbation prediction, for the four schemes defined in Eq. \eqref{eq:scheme_labels}. The fit is performed with the expression \eqref{eq:sigma_ubyu}, with $n_s=4$.
\label{fig:sigmaT_vs_u}}
\end{figure}

We present the preliminary results of the tensor running in the high-energy range (SF range) $2~\rm{GeV}\lesssim \mu\lesssim 128~\rm{GeV}$. 
We focus on \textit{u-by-u}-fits, i.e. the continuum extrapolations at fixed value of the coupling $u$:
\begin{align}
\Sigma_{\rm T}(u,a/L) &=  \sigma_{\rm T}(u) + \rho_{\rm T}(u)\left(\frac{a}{L}\right)^2 \,.
\label{eq:ubyu}
\end{align}
\begin{figure}
\includegraphics[width=0.45\textwidth]{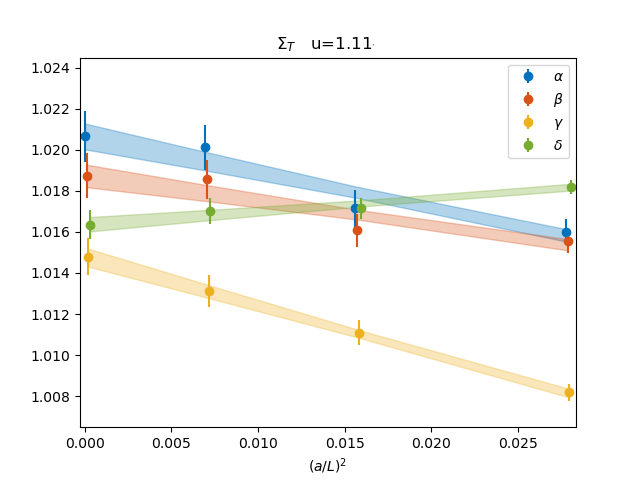} \, 
 \includegraphics[width=0.45\textwidth]{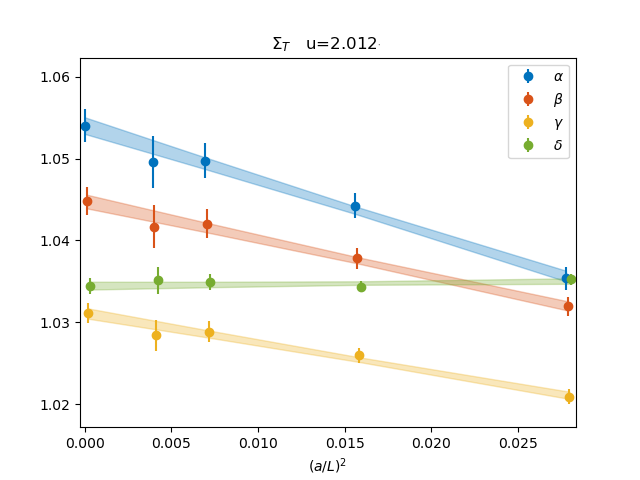}
\caption{\textit{u-by-u}-fit (Eq.\eqref{eq:ubyu}) to extract $\sigma_{\rm T}(u)$ at the lowest and highest renormalised coupling $u$ in the SF range, for the four schemes defined in Eq. \eqref{eq:scheme_labels}.
\label{fig:SigmaT}}
\end{figure}
\begin{table}
\begin{align*}
\begin{tabular}{|c|c|c|c|c|}
\hline 
$u$ & $\sigma_{T}^{\alpha}$ & $\sigma_{T}^{\beta}$ & $\sigma_{T}^{\gamma}$ & $\sigma_{T}^{\delta}$\tabularnewline
\hline 
\hline 
1.11 & 1.0206(13) & 1.0187(11) & 1.0148(09) & 1.0164(07)\tabularnewline
\hline 
1.1844 & 1.0249(13) & 1.0221(11) & 1.0166(10) & 1.0173(08)\tabularnewline
\hline 
1.2656 & 1.0243(15) & 1.0220(14) & 1.0190(12) & 1.0177(09)\tabularnewline
\hline 
1.3627 & 1.0310(19) & 1.0270(16) & 1.0209(13) & 1.0199(10)\tabularnewline
\hline 
1.4808 & 1.0345(16) & 1.0298(13) & 1.0226(10) & 1.0226(08)\tabularnewline
\hline 
1.6173 & 1.0348(23) & 1.0297(19) & 1.0252(15) & 1.0250(12)\tabularnewline
\hline 
1.7943 & 1.0434(24) & 1.0371(21) & 1.0297(16) & 1.0274(12)\tabularnewline
\hline 
2.012 & 1.0540(20) & 1.0448(17) & 1.0311(13) & 1.0345(10)\tabularnewline
\hline 
\end{tabular}
\end{align*}
\caption{$\sigma_{\rm T}(u)$ extracted from \textit{u-by-u}-fit (Eq.\eqref{eq:ubyu}), for the four schemes defined in Eq. \eqref{eq:scheme_labels}.
\label{tab:sigma}}
\end{table}
In Figure \ref{fig:SigmaT} we show the results for the lowest and highest coupling in the SF range; in Table \ref{tab:sigma} we list the values of $\sigma_{\rm T}(u)$ for all the couplings in the SF range. We observe that $\sigma_{\rm T}(u)$ tends to have smaller errors for the $\delta$-scheme. \\
The continuum $\sigma_{\rm T}(u)$ is then parametrized with two different expressions. \\
1. The first one is a polynomial in $u$:
\begin{align}
\sigma_{\rm T}(u) &=  1 +  \rho_1 u + \rho_2 u^2 + ... + \rho_{n_s} u^{n_s} \,,
\label{eq:sigma_ubyu}
\end{align}
where $\rho_1$ and $\rho_2$ are fixed by perturbation theory (ref.\cite{Pena:2017hct}):
\begin{align}
\rho_1= \gamma_0 \log2 & \, & \rho_2= \gamma_1 \log2 + \Big[\frac{1}{2} \gamma_0^2 + b_0 \gamma_0 \Big] \log2^2 \,.
\label{eq:b10-b20}
\end{align}
The result is plotted in Figure \ref{fig:sigmaT_vs_u}, with $n_s=4$. \\

\noindent 2. We also fit  $\sigma_{\rm T}(u)$  with the second expression:
\begin{align} 
\label{gamma-ubyu}
\sigma_{\rm T}(u) = \exp{\Bigl[
\int_{\sqrt{u}}^{\sqrt{\sigma(u)}} dg \frac{\gamma(g)}{\beta(g)}
\Bigr]} \,,
\end{align}
to extract directly $\gamma(g_R)$ coefficients: 
\begin{equation}
\label{eq:gamma_sum}
\gamma(g_{R})=-g_{R}^{2}\underset{n=0}{\overset{n_{t}}{\sum}}\gamma_{n}g_{R}^{2n}
\end{equation}
\begin{figure}
\centering
\includegraphics[width=0.6\textwidth]{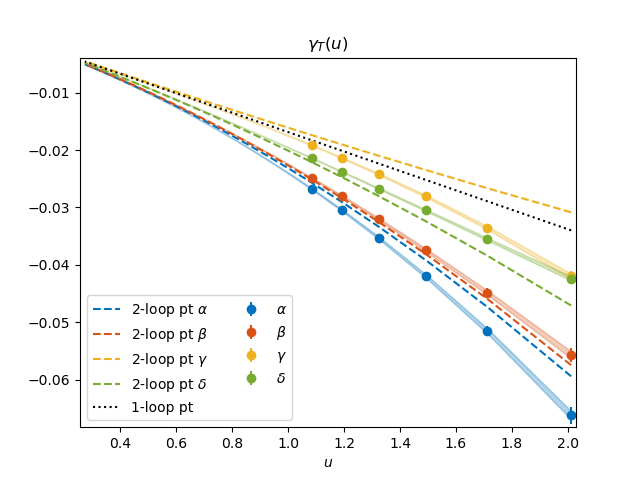}
\caption{$\gamma(u)$ compared with the universal 1-loop and the 2-loop perturbation theory, for the four schemes defined in Eq. \eqref{eq:scheme_labels}. The sum in the $\gamma$-expansion \eqref{eq:gamma_sum} stops at $n_t=2$.
\label{fig:gamma_vs_u}}
\end{figure}
The results for $\gamma(u)$ are plotted in Figure \ref{fig:gamma_vs_u}, with $n_t=2$.  We see how the non-perturbative data smoothly connect to their corresponding 2-loop predictions as the coupling $u \rightarrow 0 $. $\beta$-scheme tends to agree better with perturbation theory even at the lowest energies of the SF range. Again, $\delta$-scheme tends to have smaller errors.
The results for $\gamma(u)$ are then used to compute  the running of the tensor in the SF range:
\begin{equation}
\label{eq:Rk}
\frac{T_{R}(2^{k}\mu_{0})}{T_{R}(\mu_{0}/2)}=\exp\Big\{-\int_{g_{R}(2^{k}\mu_{0})}^{g_{R}(\mu_{0}/2)}dg\frac{\gamma(g)}{\beta(g)}\Big\}
\end{equation}
\begin{figure}
\centering
\includegraphics[width=0.6\textwidth]{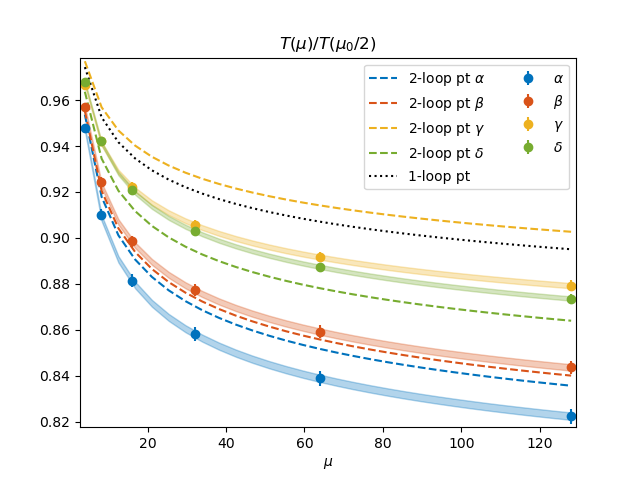}
\caption{$ T_R(2^k \mu_0)/T_R(\mu_0/2)$ compared with the universal 1-loop and the 2-loop perturbation theory,  for the four schemes defined in Eq. \eqref{eq:scheme_labels}.
\label{fig:Rk}}
\end{figure}
 The results are plotted in Figure \ref{fig:Rk}. 
 Comparing the four schemes in Table \ref{tab:Rk}, we see that $\delta$-scheme tends to have smaller errors.

\begin{table}
\begin{align*}
\begin{tabular}{|c|c|c|c|c|}
\hline 
$u$ & $\frac{(T_{R}(2^{k}\mu_{0})}{(T_{R}(\mu_{0}/2)}^{\alpha}$ & $\frac{(T_{R}(2^{k}\mu_{0})}{(T_{R}(\mu_{0}/2)}^{\beta}$ & $\frac{(T_{R}(2^{k}\mu_{0})}{(T_{R}(\mu_{0}/2)}^{\gamma}$ & $\frac{(T_{R}(2^{k}\mu_{0})}{(T_{R}(\mu_{0}/2)}^{\delta}$\tabularnewline
\hline 
\hline 
2.012 & 0,9477(18) & 0,9569(15) & 0,9668(12) & 0,9679(10)\tabularnewline
\hline 
1.7126(31) & 0,9102(25) & 0,9244(21) & 0,9421(17) & 0,9423(14)\tabularnewline
\hline 
1.4939(38) & 0,8813(29) & 0,8985(25) & 0,9223(20) & 0,9210(16)\tabularnewline
\hline 
1.3264(38) & 0,8581(31) & 0,8772(26) & 0,9058(21) & 0,9030(18)\tabularnewline
\hline 
1.1936(35) & 0,8388(32) & 0,8592(27) & 0,8916(22) & 0,8874(18)\tabularnewline
\hline 
1.0856(32) & 0,8224(32) & 0,8436(28) & 0,8792(23) & 0,8736(19)\tabularnewline
\hline 
\end{tabular}
\end{align*}
\caption{$ T_R(2^k \mu_0)/T_R(\mu_0/2)$,  for the four schemes defined in Eq. \eqref{eq:scheme_labels}.
\label{tab:Rk}}
\end{table}

\section{Conclusions}
We present the preliminary results of the non-perturbative running of the flavour non-singlet tensor operator in the high-energy range $2~\rm{GeV}\lesssim \mu\lesssim 128~\rm{GeV}$ in $\NF=3$ massless QCD, using the configuration ensembles of ref.\cite{Campos:2018ahf} and ref.\cite{DallaBrida:2016kgh}. We compare four different renormalisation schemes that differ by the normalisation of the tensor current, denoting them as $\alpha$-, $\beta$-, $\gamma$- and $\delta$-schemes.  We computed the running of the tensor bilinear and the anomalous dimension $\gamma$. At this stage of the analysis (based on \textit{u-by-u} fits), we see that errors tend to be smaller in $\delta$-scheme: e.g., for the running $ T_R(2^k \mu_0)/T_R(\mu_0/2)$, the errors are about half of those obtained in $\alpha$- or $\beta$-scheme. $\alpha$- and $\beta$-schemes correspond to the SF definitions used in refs.\cite{Pena:2017hct},\cite{Chimirri:2019xsv}). We also see that the deviations from the 2-loop predictions are smaller for the $\beta$- and $\delta$-schemes than from the $\alpha$- and $\gamma$-schemes. The observed approach of the non-perturbative data to the corresponding perturbative results defies somewhat the naive expectations one may have from the perturbative results of Eq.\eqref{eq:gamma1}. This should come as a reminder of the necessity of testing the accuracy of the available perturbative information against non-perturbative data through the study of the non-perturbative RG-running over a wide range of energy scales, reaching up to very large ones (see ref.\cite{DallaBrida:2016uha}).
We will complete the analysis at SF and  GF energy ranges.
This work is part of a long-term project which ultimately aims at providing the step scaling matrices of all four-fermion operators that contribute to $B_{\rm\scriptscriptstyle{K}}$ in the Standard Model and beyond, as outlined in ref.\cite{Campos:2019nus}.

\acknowledgments

We wish to thank Alessandro Conigli, Patrick Fritzsch, Carlos Pena, David Preti, Alberto Ramos and Pol Vilaseca for their help. 
This work  is partially supported by INFN and CINECA, as part of research project of the QCDLAT INFN-initiative. 
We acknowledge the Santander Supercomputacion support group at the University of Cantabria which provided access to the Altamira Supercomputer at the Institute of Physics of Cantabria (IFCA-CSIC).
We also acknowledge support by the Poznan Supercomputing and Networking Center (PSNC) under the project with grant number 466. AL acknowledges support by the U.S.\ Department of Energy under grant number DE-SC0015655.

\end{document}

%% file: macros.tex
\newcommand{\ba}{\begin{array}}
\newcommand{\ea}{\end{array}}

\newcommand{\Dslash}{\relax{\kern+.25em / \kern-.70em D}}

\newcommand{\Real}{\relax{\mathsf{\Gamma\kern-.35em R}}}
\newcommand{\Int}{\relax{\mathsf{Z\kern-.40em Z}}}

\newcommand{\NF}{N_\mathrm{\scriptstyle f}}

\newcommand{\gbar}{\kern1pt\overline{\kern-1pt g\kern-0pt}\kern1pt}
\newcommand{\mbar}{\kern2pt\overline{\kern-1pt m\kern-1pt}\kern1pt}
\newcommand{\obar}[1]{\kern3pt\overline{\kern-2pt #1\kern-0pt}\kern1pt}

\newcommand{\fA}{f_{\rm\scriptscriptstyle A}}

\newcommand{\kV}{k_{\rm\scriptscriptstyle V}}
\newcommand{\kT}{k_{\rm\scriptscriptstyle T}}

\newcommand{\gV}{g_{\rm\scriptscriptstyle V}}

\newcommand{\lV}{l_{\rm\scriptscriptstyle V}}
\newcommand{\lT}{l_{\rm\scriptscriptstyle T}}

\newcommand{\ZT}{Z_{\rm\scriptscriptstyle T}}

\newcommand{\abar}{\kern1pt\overline{\kern-1pt a\kern-0.5pt}\kern1pt}

